\documentclass[10pt, conference, letterpaper]{IEEEtran}
\usepackage{amsmath}
\usepackage{tabularx,multirow}
\usepackage{graphicx,subfigure,comment}
\usepackage[table]{xcolor}
\usepackage{url} 
\usepackage{psfrag,dsfont}
\usepackage{cite}
\usepackage{siunitx}
\usepackage{tabularx}

\usepackage[algoruled,medskip,dontprintsemicolon,linesnumbered,Algorithm]{algorithm}
\usepackage[noend]{algpseudocode}

\newcommand{\Fig}[1]{Fig.~\ref{fig:#1}}
\newcommand{\Sec}[1]{Sec.~\ref{sec:#1}}
\newcommand{\Tab}[1]{Tab.~\ref{tab:#1}}
\newcommand{\Eq}[1]{(\ref{eq:#1})}

\newcommand{\ind}[1]{\mathds{1}_{\left[#1\right]}}

\begin{document}

\title{
5G Traffic Forecasting:\\
If Verticals and Mobile Operators Cooperate
} 

\author{\IEEEauthorblockN{Francesco Malandrino}
\IEEEauthorblockA{CNR-IEIIT, Politecnico di Torino, CNIT\\
Email: francesco.malandrino@ieiit.cnr.it}
\and
\IEEEauthorblockN{Carla-Fabiana Chiasserini}
\IEEEauthorblockA{Politecnico di Torino, CNR-IEIIT, CNIT\\
Email: chiasserini@polito.it}
}

\maketitle

\begin{abstract}
In 5G research, it is traditionally assumed that vertical industries (a.k.a verticals) set the performance requirements for the services they want to offer to mobile users, and the mobile operators alone are in charge of {\em orchestrating} their resources so as to meet such requirements. Motivated by the observation that successful orchestration requires reliable traffic predictions, in this paper we investigate the effects of having the verticals, instead of the mobile operators, performing such predictions. Leveraging a real-world, large-scale, crowd-sourced trace, we find that involving the verticals in the prediction process reduces the prediction errors and improves the quality of the resulting orchestration decisions.
\end{abstract}

\section{Introduction and related work}

Unlike their fourth-generation counterparts, 5G networks will not only transport data, but also process them. Network, computing, and memory resources controlled by mobile network operators (MNOs), will concurrently support multiple services, under the {\em network slicing} paradigm~\cite{slicing1,slicing2}. It is universally expected~\cite{slicing1,slicing2,slicing-survey} that vertical industries (e.g., automotive or media companies) specify the {\em requirements} of their services, i.e., which computation must be performed and the associated target key performance indicators (KPIs). MNOs, on the other hand, have to manage their network so as to ensure that all target KPIs are met at the lowest cost for themselves, a problem known as service orchestration~\cite{slicingalgos,orch-arch}.

Our purpose in this paper is to study a different model of interaction between vertical industries (henceforth verticals) and MNOs, whereby verticals provide not only the target KPIs but also an estimation of their expected traffic patterns. The reason for this change is that service orchestration is greatly simplified if the evolution of the demand to serve is known~\cite{noi-infocom18}, or it can be reliably predicted~\cite{scianca}, and verticals are in a better position than MNOs to make such a prediction. 
Indeed, unlike verticals, MNOs cannot access, for technical and legal reasons, detailed, application-layer information on the  traffic  flowing through their network. It follows that, since 
network slices are tailored around a single type of service\footnote{I.e., services with the same KPIs.}, the service-specific predictions that verticals can make may be more useful than  predictions made by MNOs.

Our first task is therefore to compare the accuracy of the predictions that MNOs and verticals can make based on the information they can access. To this end, we leverage a real-world, large-scale, crowd-sourced trace, containing mobility and traffic information about over 90,000 users in the Los Angeles area. Thanks to its crowd-sourced nature, the trace contains a superset of the information available to verticals and MNOs; therefore, we can consider a state-of-the-art prediction technique, feed it the data available to MNOs and verticals, and check which of them yields the most accurate result.

Beyond the accuracy of predictions, we are interested in the effect of prediction errors on the resulting orchestration decisions. Specifically, we are interested in two adverse consequences of inaccurate predictions, namely, (i) unused capacity, when the demand is overestimated and the network slice is over-provisioned, and (ii) scale-up events, when the capabilities of an under-provisioned slice must be swiftly improved to face a higher-than-predicted demand. Our task is to establish which of these events is more common and how the predictions obtained by verticals and MNOs affect them.

Finally, we compare both alternatives against a scenario where verticals and MNOs share not only the traffic prediction but the input information they use to make them. This serves as a useful benchmark,a lthough it would be realistic only in very specific scenarios, e.g., when the 
MNO also acts as a  vertical and provides  value-added services such as video calls or streaming. 

The rest of this paper is organized as follows. \Sec{dataset} presents the real-world dataset we use for the forecasts, while \Sec{algos} describes the techniques we adopt. After presenting the metrics of interest and numerical results in \Sec{results}, we conclude the paper in \Sec{conclusion}.

\section{A real-world dataset}
\label{sec:dataset}

For our analysis, we use a real-world, {\em crowd-sourced} dataset collected from the WeFi app~\cite{wefi}. WeFi provides its users with location-specific information on the available Wi-Fi networks, and such information is crowd-sourced from the users themselves. Specifically, the app tracks:
\begin{itemize}
    \item the current time (with a one-hour granularity) and location;
    \item the mobile operator and cell the user is associated with (if any);
    \item the SSID and BSSID of the Wi-Fi network she is connected to (if any);
    \item the amount of data (uplink and downlink) used by the currently-active application, and the identity of the application itself.
\end{itemize}
New records in the trace are created every time any of the above pieces of information changes, e.g., the user switches between apps. The features of the trace are summarized in \Tab{trace}.

\begin{figure}[t]
\centering
\includegraphics[width=1\columnwidth]{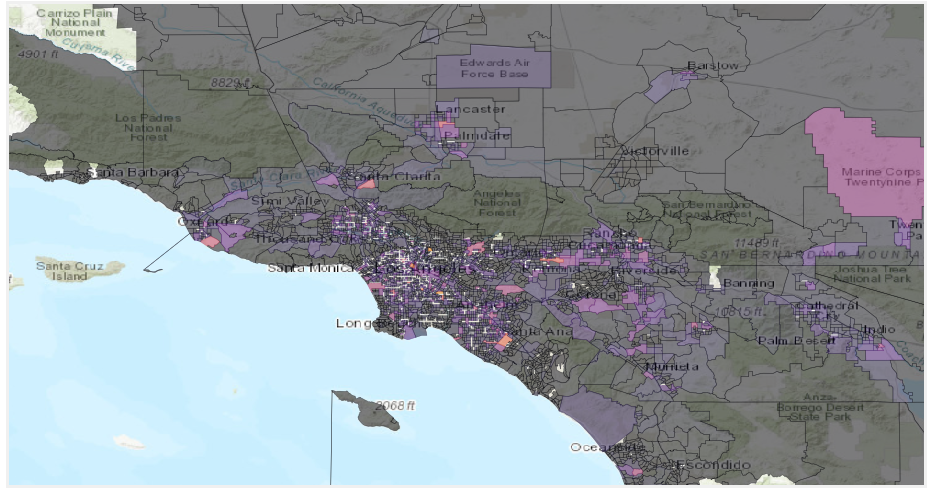}
\caption{The area covered by the WeFi dataset. Colors reflect the average download rate considering all applications; warmer colors correspond to a higher traffic demand.
    \label{fig:map}
} 
\end{figure}

\Fig{map} depicts the area covered by the trace -- greater Los Angeles -- and the traffic density therein. We can observe a higher traffic demand in the most densely-populated zones, e.g., downtown Los Angeles, and a lower demand in rural or wilderness areas. Also notice, in the far East and North of the map, the Edwards and Twentynine Palms military bases, with a much higher traffic than the surrounding rural areas.

\begin{table}[t]
\caption{
\label{tab:trace}
The WeFi trace.
} 
\begin{tabularx}{.9\columnwidth}{|c|X|}
\hline
Metric & Value\\
\hline
\hline
Covered area & $98\times 79\text{km}^2$ \\
\hline
Collection time & March 2016 \\
\hline
Number of records & 835 million \\
\hline
Unique users & 91,341 \\
\hline
Unique cells & 1,552 \\
\hline
Unique BSSIDs & 457,388 \\
\hline
Unique apps & 72,958 \\
\hline
Total traffic & 535 TByte \\
\hline
Coverage & 2\% (WeFi estimate) \\
\hline
\end{tabularx}
\end{table}

Importantly, unlike similar datasets collected by mobile operators~\cite{trace-old,sharing}, the WeFi trace includes information on different mobile operators and technologies (including Wi-Fi), as well as different applications -- an aspect that makes the trace especially well-suited to study 5G networks. Indeed, the network slicing paradigm is predicated on tailoring slices to individual applications; in this context, knowledge on application-specific traffic patterns is much more useful than information on global demand fluctuations.

\section{Forecasting technique}
\label{sec:algos}

Here we briefly describe the data available to MNOs and to verticals, the prediction technique that we apply, and the output (i.e., the predictions) that MNOs and verticals can obtain.

{\bf Input and output data.}
The information available to MNOs, i.e., in machine learning jargon, the {\em features} their forecast is based upon, include, for each cell and time period:
\begin{itemize}
    \item the total demand (by all users, for all apps);
    \item the number of users covered by the cell (regardless their activity).
\end{itemize}
In the case of the WeFi trace, time periods correspond to one-hour time intervals. Notice that, due to technical and legal reasons, MNOs have no knowledge of what individual users do, i.e., which app(s) they use.

Verticals, on the other hand, only have information about their own service\footnote{For simplicity, we assume that a service corresponds to an app, and that each vertical only provides one service.}. On the positive side, they know the identity of their users and their fine-grained location -- in the case of the WeFi trace, a $10\times 10~\text{m}^2$~{\em tile}. For each time period and tile, verticals can thus keep track of:
\begin{itemize}
    \item the traffic demand for their service;
    \item the number of users of their service;
    \item their traffic history, e.g., the amount of data they downloaded in the past.
\end{itemize}

For both MNOs and verticals, the quantity to predict is the {\em total} demand of a specific app, i.e., the traffic the network slice will process. As it is commonplace in machine learning, we split our dataset into a training set, including the first three weeks of the trace, and a testing set, containing the last one.

{\bf Prediction technique.}
The prediction task under study belongs to the class of {\em time-series forecasting} problems. Input data are multi-variate, i.e., the forecast must be based on the evolution of multiple quantities -- in the case of MNOs, for example, the traffic demand of each individual cell. This rules out the use of traditional approaches like the Holt-Winters method adopted in~\cite{scianca}, which requires uni-variate time series.

We therefore turn to neural networks, namely, Long Short-Term Memory (LSTM) networks, introduced in~\cite{lstm} and, since then, successfully applied in a variety of fields, from computer vision to voice recognition. Specifically, we use the implementation from Google's TensorFlow~\cite{tensorflow} library, accessed through the Keras~\cite{keras} high-level front-end.

\section{Numerical results}
\label{sec:results}

This section shows the predicted behavior of the traffic demand, describes the performance metrics we consider (\Sec{perf}), and discusses the results we obtain (\Sec{results-results}).

\begin{figure*}
\centering
\subfigure[\label{fig:traffic-yt}]{
    \includegraphics[width=.3\textwidth]{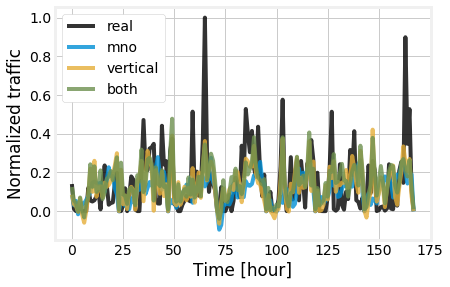}
} 
\subfigure[\label{fig:traffic-fb}]{
    \includegraphics[width=.3\textwidth]{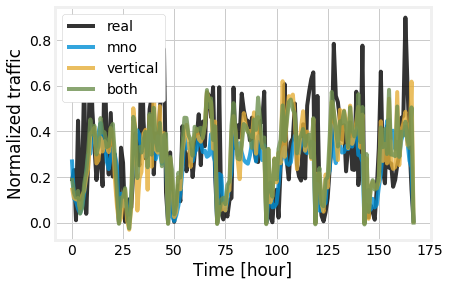}
} 
\subfigure[\label{fig:traffic-nf}]{
    \includegraphics[width=.3\textwidth]{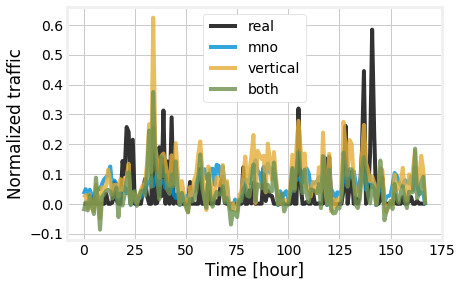}
} 
\caption{Actual and predicted traffic for YouTube (a), Facebook (b), and Netflix (c).\label{fig:traffics}} 
\end{figure*}

\begin{figure}
\centering
\includegraphics[width=.4\textwidth]{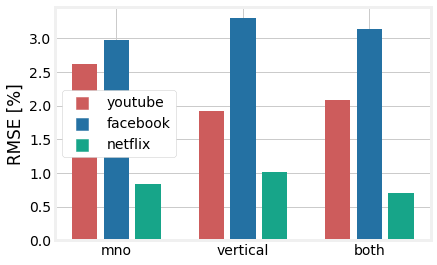}
\caption{Prediction error (RMSE) for different apps and scenarios.\label{fig:rmse}} 
\end{figure}

\begin{figure}
\centering
\includegraphics[width=.4\textwidth]{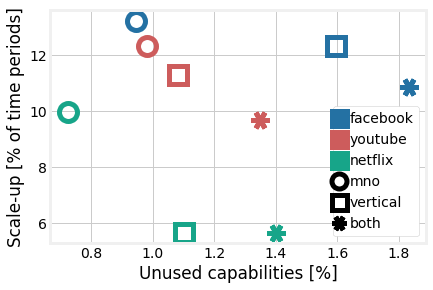}
\caption{Unused capabilities and scale-up events for different apps and scenarios.\label{fig:performance}} 
\end{figure}

\Fig{traffics} shows the actual traffic (black line), as well as the traffic predicted:
\begin{itemize}
    \item using the data available to MNOs (blue lines);
    \item using the data available to verticals (yellow lines);
    \item using both (green lines).
\end{itemize}

\subsection{Performance metrics}
\label{sec:perf}

As mentioned in the introduction, we are interested in two main aspects, namely, the prediction accuracy and the quality of orchestration decisions. 
Quantifying the prediction accuracy is fairly straightforward; specifically, we resort to the well-known RMSE (reduced mean-square error) metric, defined as:
\begin{equation}
\label{eq:rmse}
\text{RMSE}=\sqrt{\frac{1}{n}\sum_{j=1}^n\left(y_j-\hat{y}_j\right)^2},
\end{equation}
where~$n$ is the number of time periods the forecast extends across (seven days, i.e., 168~one-hour periods in our case), while~$y_j$ and~$\hat{y}_j$ are, respectively, the actual and predicted traffic at the $j^\text{th}$ time period. Note that all metrics are computed separately for each app.

With regard to orchestration decisions, there are two adverse effects we seek to minimize. The first is unused capabilities, i.e., network slices that are over-provisioned with respect to the actual traffic demand. We can quantify unused capabilities as:
\begin{equation}
\label{eq:waste}
w=\sum_{j=1}^n\max\left(0,\hat{y}_j-y_j\right).
\end{equation}
That is, for each time period, we consider the difference between the predicted demand (according to which the slice is dimensioned) and the actual one; if positive, such a difference is representative of the amount of unused capabilities.

The second adverse effect is represented by scale-up events, i.e., under-provisioned slices whose capabilities have to be swiftly extended to cope with unforeseen increases in traffic. As discussed in~\cite{ztorch}, such events have the potential to decrease the QoS/QoE of all services supported by the MNO. The quantity~$u$ expresses the number of time periods in which such events happen:
\begin{equation}
\label{eq:scaleup}
u=\sum_{j=1}^n\ind{y_j>\hat{y}_j}.
\end{equation}

\subsection{Results}
\label{sec:results-results}

Each of the plots in \Fig{traffics} refers to one of the three most used apps in the trace: YouTube, Facebook, and Netflix. 
As one might expect, the traffic demand exhibits clear weekly and daily patterns, e.g., morning and evening peaks. However, the magnitude of traffic peaks is not consistent throughout all days, e.g., see the peaks around periods~60 and~160 in \Fig{traffic-yt}. This feature of traffic patterns makes prediction harder; indeed, we can observe that these higher-than-usual peaks are never properly predicted, regardless of the scenario.

\Fig{rmse} shows the prediction error (RMSE, as defined in \Eq{rmse}) associated with different services and scenarios. A first fact to notice is that the prediction error is, in general, fairly small: a testament to the effectiveness of the LSTM prediction technique, as well as to the overall regularity of the traffic demand. It is perhaps even more interesting to observe that the RMSE changes significantly across apps; comparing \Fig{rmse} to  \Fig{traffics}, we can conclude that more numerous and irregular peaks are associated with a higher prediction error.

The effect of shifting the task of traffic prediction from the MNO (first group of bars in \Fig{rmse}) to the vertical (second group of bars) is also inconsistent across services. The vertical seems to be better than the MNO at predicting YouTube, but the opposite is true for Facebook; as for Netflix, no significant difference can be observed. The third group of bars refers to the benchmark scenario where MNOs and verticals share their information and jointly predict the demand: in this case, the resulting error is close to the lowest of the errors yielded by the other two scenarios.

\Fig{performance} focusses on the impact of the traffic prediction on orchestration decisions. The $x$-axis therein shows the total unused capabilities, i.e., the $w$-metric defined in \Eq{waste}, while the $y$-axis shows the frequency of scale-up events, i.e., the $u$-metric defined in \Eq{scaleup}. Each dot corresponds to a combination of app (identified by its color) and scenario (identified by the marker used), e.g., the red square corresponds to the vertical making predictions about YouTube traffic.

A first observation we can make is that scale-up events are much more common than unused capabilities. Indeed, scale-up events happen in around 10\% of time periods (i.e., roughly twice per day), while unused capabilities account for less than 2\% of the total. This is unwelcome news, since scale-up events are, in most real-world cases, a more serious issue than unused capabilities, and far more likely to result in a violation of target KPIs and/or higher costs for the MNO~\cite{ztorch}. This seems to disagree with the low RMSE values summarized in \Fig{rmse}; however, it is worth recalling that {\em any} underestimation of the demand, no matter how slight, leads to a scale-up event. For the same reason, the Netflix app (green markers) is associated with almost the same number of scale-up events as Facebook and YouTube, in spite of the much lower RMSE.

Even more interestingly, there is a remarkably consistent relationship between the different scenarios. MNO predictions yield the highest number of scale-up events and the lowest amount of unused capabilities; moving to vertical and joint predictions has the effect of reducing the scale-up events in exchange for a small increase in unused capabilities. It is important to remark how this happens for all apps, in spite of the different levels of prediction accuracy (\Fig{rmse}) they are associated with. In other words, while involving the verticals in traffic prediction does not necessarily improve the predictions accuracy {\em per se}, it does yield better orchestration decisions, with a healthier balance between scale-up events and unused resources.

\section{Conclusion and Future Work}
\label{sec:conclusion}

We considered the orchestration problem in 5G networks based on network slicing. After remarking that good orchestration decisions depend on accurate traffic predictions, we investigated whether verticals are in a better position than MNOs to make such predictions. Using a real-world, large-scale, crowd-sourced trace, we found that, while the prediction error is not consistent throughout different apps, involving the verticals in the prediction leads to a slightly higher amount of unused capacity and, more importantly, to a lower frequency of scale-up events.

A first direction to extend our work is considering additional prediction techniques, including generalizations of the Holt-Winters method. Furthermore, we can couple the predictions with actual orchestration algorithms, including state-of-the-art approaches taken from the literature as well as purpose-built ones.

\bibliographystyle{IEEEtran}
\bibliography{refs}

\section*{Acknowledgment}

This work is supported by the European Commission through the H2020 projects 5G-TRANSFORMER (Project ID 761536) and 5G-EVE (Project ID 815074).

\end{document}